\documentclass[aps,pra,nofootinbib,twocolumn]{revtex4}

\usepackage{amsmath, amsthm, amssymb,graphicx,color,bbm}
\usepackage{courier}
\usepackage{textcomp}
\usepackage[urlcolor=blue]{hyperref}
\usepackage[utf8]{inputenc}

\DeclareFixedFont{\ttb}{T1}{txtt}{bx}{n}{12} 
\DeclareFixedFont{\ttm}{T1}{txtt}{m}{n}{12}  

\usepackage{color}
\definecolor{deepblue}{rgb}{0,0,0.5}
\definecolor{deepred}{rgb}{0.6,0,0}
\definecolor{deepgreen}{rgb}{0,0.5,0}
\definecolor{keywords}{RGB}{255,0,90}
\definecolor{comments}{RGB}{0,0,113}

\usepackage{listings}

\newcommand\pythonstyle{\lstset{
language=Python,
basicstyle=\ttfamily\small,
otherkeywords={self},             
keywordstyle=\bfseries\color{deepblue}\small,
commentstyle=\color{comments}\small,
emph={MyClass,__init__},          
emphstyle=\bfseries\color{deepred}\small,    
stringstyle=\color{deepgreen}\small,
frame=single,                     
framesep=1em,
showstringspaces=false,            %
upquote=true
}}

\lstnewenvironment{python}[1][]
{
\pythonstyle
\lstset{#1}
}
{}

\newcommand\pythoninline[1]{{\pythonstyle\lstinline!#1!}}

\begin{document}

\title{Probing quantum processor performance with pyGSTi}
\begin{abstract}
PyGSTi is a Python software package for assessing and characterizing the performance of quantum computing processors.  It can be used as a standalone application, or as a library, to perform a wide variety of quantum characterization, verification, and validation (QCVV) protocols on as-built quantum processors.  We outline pyGSTi’s structure, and what it can do, using multiple examples. We cover its main characterization protocols with end-to-end implementations.  These include gate set tomography, randomized benchmarking on one or many qubits, and several specialized techniques. We also discuss and demonstrate how power users can customize pyGSTi and leverage its components to create specialized QCVV protocols and solve user-specific problems.
\end{abstract}

\author{Erik Nielsen}
\author{Kenneth Rudinger}
\author{Timothy Proctor}
\author{Antonio Russo}
\author{Kevin Young}
\author{Robin Blume-Kohout}

\date{\today}
\maketitle

\tableofcontents

\section{What is pyGSTi?}
PyGSTi\cite{pyGSTi0.9.9.1} is software to probe, analyze, and assess the performance of digital quantum computing (QC) processors containing 1-20$^+$ qubits.  PyGSTi can be used to:
\begin{enumerate}
\item Construct QC experiments (collections of quantum circuits) designed to probe many different aspects of a quantum computer's behavior.
\item Specify those experiments precisely in human- and machine-readable formats.
\item Read, store, and manipulate the data produced by running those experiments.
\item Analyze that data to estimate key performance parameters (e.g., error rates) and create models for the processor (e.g. process matrices and gate sets).
\item Display, visualize, and interactively explore that analysis.
\end{enumerate}
PyGSTi does \emph{not} currently
\begin{enumerate}
\item compare QC processors against their classical competitors,
\item benchmark non-digital QC processors (e.g. adiabatic QC, quantum annealers, or analogue simulators), or
\item compare different implementations of the same quantum algorithm (reflecting, e.g., different compilers or qubit connectivities)
\end{enumerate}
Instead, pyGSTi focuses on testing \emph{how well a digital quantum computer lives up to its specification}.  PyGSTi's protocols, algorithms, and supporting routines enable users to check whether a processor is behaving according to spec, identify components that are \emph{not} close to spec, and reveal exactly \emph{how} those components are failing.  These capabilities let pyGSTi users understand the strength and nature of  unwanted noise in as-built QC processors.  They can be used to tune and ``debug'' existing hardware, and to enable improvements in the next generation of processors.  

PyGSTi contains a lot of independent modules that support these core goals, and can be useful for solving other, related QC problems.  We mention some of those applications throughout this paper.  PyGSTi is a medium-sized Python package that can be used as a standalone application or a library. At the time of this writing, the latest version of pyGSTi is 0.9.9.1.  It contains around 150,000 lines of code, and is distributed under the open-source Apache version 2 license.  It started life (circa 2013) as research code, but a substantial effort has been invested in making pyGSTi user-friendly and accessible to professionals in QC.  The central aim of this paper is to describe the things that pyGSTi can do.

Readers unfamiliar with quantum computing and/or techniques for characterizing QC performance should read Section \ref{secBackground}.  Expert readers may wish to skip directly to the discussion of pyGSTi's capabilities and use cases in Section \ref{secCapabilities}.

\section{Background\label{secBackground}}

``Quantum software'' can mean many things, including both quantum algorithms that run on quantum computers \emph{and} traditional software for classical computers that enables the use or development of QC technology.  PyGSTi is the latter.  

Classical software can support QC development in many ways.  PyGSTi is designed to help with \emph{quantum processor characterization}, also known as ``quantum characterization, verification, and validation'' or \emph{QCVV}.  This means measuring the differences between an as-built quantum processor's behavior and that of an ideal processor.  The purpose of a digital quantum computer is to run programs, which are described by quantum circuits.  So to characterize a processor, we run circuits and study the results.  

The objectives of quantum processor characterization are:
\begin{enumerate}
\item to quantify how well a QC processor is performing, often by creating an \emph{error model} for it,
\item to predict its performance on arbitrary circuits that might be run in the future,
\item to extrapolate how this hardware would be expected to perform in other contexts (e.g. with more qubits or better control).
\item to guide development of next-generation hardware by understanding the types and magnitudes of errors that appear in current processors.
\end{enumerate}
Tools to achieve these objectives are \emph{QCVV protocols}.  These are specific recipes for measuring some aspect of a QC processor's performance.  Each protocol combines (1) a list of circuits to run on the processor being tested, and (2) an algorithm to analyze the resulting data.  The QCVV toolbox contains many complementary protocols (with more introduced regularly) to assess different aspects of QC performance.

We find it useful to divide existing protocols into three groups:
\begin{enumerate}
\item Tomographic protocols attempt to fit a detailed error model.  They include quantum state and process tomography\cite{Hradil1997qst,chuangnielsen1996qpt,knee2018qpt}, gate set tomography (GST) \cite{blume2016certifying,dehollain2016optimization,merkel2013self}, randomized benchmarking tomography \cite{kimmel2014robust}, robust phase estimation (RPE) \cite{kimmel2015robust}, Hamiltonian estimation/tomography \cite{schirmer2004hamiltoniantomograhpy}, and reduced-model ``compressed sensing'' techniques \cite{donoho2006compressed,cramer2010mpstomography}.
\item Coarse-grained benchmarking/metrology protocols measure specific error rates but do not attempt to model every aspect of behavior.  They include direct fidelity estimation \cite{flammia2011directfidelity}, randomized benchmarking (RB) \cite{magesan2011scalable,knill2008randomized,proctor2018direct}, cross-entropy benchmarking \cite{boixo2018characterizing}, and cycle benchmarking \cite{erhard2019characterizing}.
\item Holistic benchmarking measures overall performance without specifically estimating gate error rates.  These include IBM's protocol to estimate quantum volume \cite{cross2018validating}, volumetric benchmarks \cite{blume2019volumetric}, and small instances of quantum algorithms \cite{childs2018speedup}.
\end{enumerate}

QCVV protocols have been used extensively by QC researchers to characterize and debug many qubits and testbed processors.  We do not have the resources to provide a comprehensive bibliography here!  PyGSTi implements many, but not all, of these tools.  Some examples of experimental QCVV (emphasizing experiments that used pyGSTi) include Refs \cite{knill2008randomized,gambetta2012characterization,kim2015dotqubit,fogarty2015nonexponential,dehollain2016gst,blume2016certifying,cross2017open,rudinger2017experimental,mckay2017three,proctor2018direct,rudinger2018probing,boixo2018supremacy,proctor2019detecting,erhard2019cyclebenchmarking,white2019performance}.


\section{pyGSTi capabilities\label{secCapabilities}}  
PyGSTi helps users to perform QCVV protocols.  It can generate QCVV experiments, translate them into machine-readable form for running on real QC hardware and/or simulate them with user-specified noise models, analyze the data from QCVV experiments, and (finally) help to visualize and summarize the results.  Many different QCVV protocols are implemented in pyGSTi.  They fall into three clusters:

\begin{enumerate}
\item \textbf{Gate set tomography (GST).} PyGSTi began as a GST implementation.  Today, it implements several kinds of GST.  GST protocols  \cite{blume2016certifying,dehollain2016optimization,merkel2013self} fit detailed gate-level error models to the results of running a set of structured, periodic circuits of varying lengths.  Variants either (1) fit different models, and/or (2) run different circuits.  GST models generally represent each operation in the processor's API with some kind of quantum process matrix, so they are \emph{Markovian} models (although GST can be extended to specific forms of non-Markovian noise).  If GST can find a  model that fits the data, it usually describes the processor's errors completely.  GST models tend to have a lot of parameters, so GST requires running many circuits and is considered relatively resource-intensive.  GST protocols in pyGSTi include:
\begin{itemize}
\item Linear GST
\item Long-sequence (standard) GST
\item Streamlined GST (much smaller experiment size)
\item Reduced-model GST (smaller, special-purpose models)
\end{itemize}
\item \textbf{Randomized benchmarking (RB).} RB protocols \cite{magesan2011scalable,knill2008randomized,proctor2018direct} define ensembles of random circuits whose \emph{average} success probability decays exponentially at a rate $r$ that can be interpreted as an average ``error rate'' for the gates in those circuits.  RB protocols define and measure just a few error metrics, and generally demand less data (fewer circuits) than GST.  Their simplicity makes data analysis easy (just fit an exponential decay) and grants some robustness against contamination by unexpected errors.  RB does one thing -- extract a single aggregated score for the overall goodness of a set of qubits -- and does it well.  PyGSTi implements several RB protocols including:
\begin{itemize}
\item Clifford (standard) RB \cite{magesan2011scalable}, suitable for 1-3 qubits
\item Direct RB \cite{proctor2018direct}, suitable for up to 10 qubits
\item Mirror RB, scalable to $>10$ qubits
\item Simultaneous RB \cite{gambetta2012characterization}, used in conjunction with any type of RB to detect certain kinds of crosstalk
\end{itemize}
\item \textbf{Special-purpose protocols.}  PyGSTi supports a number of more specialized characterization protocols for a variety of tasks.  These include (but are not limited to):
\begin{itemize}
\item \textbf{Robust phase estimation (RPE).} RPE \cite{kimmel2015robust} measures the rotation angle[s] (i.e., phases) of a gate, correcting for most forms of decoherence and noise.
\item \textbf{Dataset comparison.}\cite{rudinger2018probing}  This simple but effective protocol identifies statistically significant changes between two datasets that \emph{should} be identical (up to statistical fluctuations), to detect drift, crosstalk, or other forms of context dependence.
\item \textbf{Drift tracking.}\cite{proctor2019detecting}  This protocol analyzes time-stamped results from repeating one or more circuits repeatedly over time, looks for evidence of drifting behavior, and reconstructs time trajectories of the drifting parameters.
\item \textbf{Volumetric benchmarking.}\cite{blume2019volumetric}  This family of protocols runs relatively large circuits of various sizes with a common structure and collates the results to delineate the processor's overall capability to run circuits.
\end{itemize}
\end{enumerate}


Each of the examples given above can be run by a pyGSTi user ``out of the box,'' with minimal coding, from a Python shell (e.g. Jupyter).  User-friendly tutorials are available at \texttt{http://www.pygsti.info}.  However, pyGSTi is \emph{also} a library, and users are also free (and encouraged) to use it to re-engineer or build QCVV protocols from scratch.  A few of the example protocols listed above could feasibly be implemented from scratch by a student or scientist (e.g., RB has been implemented in many labs), but most of the protocol implementations in pyGSTi (especially GST) represent months or years of effort, and most users will not wish to reimplement them.  Section \ref{secUseCases} delves into specific applications in more detail, and also illustrates how some of pyGSTi's supporting routines provide useful applications in their own right.

A brief remark is in order about circuit simulation.  Other software packages exist for which forward simulation is a \emph{primary} focus (e.g., QuTip\cite{johansson2013qutip}, quantumsim\cite{quantumsim}, LIQUi$|>$\cite{wecker2014liqui}).  PyGSTi is not in any real sense a competitor to these packages.  PyGSTi's internal circuit simulation is uniquely optimized for specific QCVV use cases (e.g., very fast strong simulation of highly periodic circuits on a few qubits, and frequent re-evaluation of those results with slightly different noise models).  This capability may be useful to users.  However, we (the pyGSTi authors) would like to see pyGSTi interface more closely with dedicated circuit simulation packages that might have superior performance.

\section{pyGSTi basics\label{secBasics}}
Before delving into use cases, we give a brief overview of pyGSTi's structure -- its major components, and how they interact.  This material is a little dry, but important for making sense of the code snippets in the next section on use cases.  More details about pyGSTi's structure can be found in the \href{https://github.com/pyGSTio/pyGSTi/tree/9d516b0b7edf6ff03e75eb148eb58f25c2f34604/jupyter_notebooks}{many tutorial notebooks included in the pyGSTi repository}.

PyGSTi is like New Mexican cuisine: it combines just a few ingredients in many different ways (to make lots of good stuff).  PyGSTi is built around three main Python objects:
\begin{itemize}

\item \textbf{Circuits}: the \texttt{pygsti.objects.Circuit} Python class represents a sequence of quantum operations on qubits/qudits.  Circuits are similar to Python tuples of circuit layers, and circuit layers are similar to tuples of gate labels.  (A ``layer'' is a collection of operations that are executed concurrently.)  Circuits can be translated to IBM's OpenQASM language \cite{cross2017open} and Rigetti Computing's Quil language \cite{smith2016practical}, allowing characterization protocols to easily be run on widely accessible ``cloud'' devices \cite{ibmq2, rigetti-qcs}.
\item \textbf{Models}: the \texttt{pygsti.objects.Model} Python class encapsulates a model for a noisy quantum processor.  The canonical example is a \emph{gate set}, but pyGSTi \texttt{Model}s are highly flexible.  Models have zero or more adjustable \emph{parameters}.  The chief role of a model is to simulate circuits, so any \texttt{Model} object can take a \texttt{Circuit} object and generate its predicted outcome probabilities.  The most commonly used models are gate sets on 1-2 qubits.  These behave like dictionaries of process matrices, indexed by a gate label.  For instance, \texttt{myModel[('Gxpi2',0)]} might be a $4\times 4$ Numpy \cite{numpy} array giving the Pauli transfer matrix for a $\pi/2$ X-rotation gate on qubit 0.

\item \textbf{Data sets}: the \texttt{pygsti.objects.DataSet} Python class holds the outcome counts for a set of circuits.  A \texttt{DataSet} behaves like a Python dictionary whose keys are \texttt{Circuit} objects and whose values are dictionaries of outcome counts.  For example, a 2-qubit data set might associate the dictionary \texttt{\{'00': 10, '01': 55, '10': 5, '11': 30\}} of outcome counts with the circuit \texttt{Circuit([Gxpi2:0][Gypi2:1])}.
\end{itemize}

More detailed descriptions of these objects are found in the pyGSTi tutorial on \href{https://github.com/pyGSTio/pyGSTi/blob/9d516b0b7edf6ff03e75eb148eb58f25c2f34604/jupyter_notebooks/Tutorials/01-Essential-Objects.ipynb}{essential objects}.  Most other objects that appear in pyGSTi code are just containers for these three primary objects.  For example, a \texttt{MultiDataSet} is a collection of data sets, a \texttt{ProcessorSpec} is a collection of models (with additional meta information), and a \texttt{ModelEstimateResults} object contains datasets, circuits, and models.  In the analogy to New Mexican cuisine, if \texttt{Circuit}, \texttt{Model}, and \texttt{DataSet} are like beans, cheese, and meat, then these other objects are tacos, enchiladas, and burritos.

Most of pyGSTi's \emph{algorithms} use the three primary object types above as their input and output (sometimes wrapped in containers).  Common pyGSTi tasks combine and map between them:
\begin{itemize}
\item \textbf{Circuit simulation}:

  \texttt{Circuit} + \texttt{Model} $\implies$ probabilities (floats)
\item \textbf{Simulated outcome counts}:

  \texttt{Circuit}s +  \texttt{Model} $\implies$ \texttt{DataSet}
\item \textbf{Model testing}:

  \texttt{Circuit}s +  \texttt{Model} =?= \texttt{DataSet}
\item \textbf{GST}: \texttt{Circuit}s + \texttt{DataSet} $\implies$  \texttt{Model}
\item \textbf{RB}: \texttt{Circuit}s + \texttt{DataSet} $\implies$ Error rate (float)
\end{itemize}

\begin{widetext}

\section{pyGSTi applications and use cases\label{secUseCases}}
We have seen pyGSTi used in three distinct ``use modes'':
\begin{enumerate}
\item As a standalone QCVV tool in an interactive environment,
\item As a Python library, called by user-written code,
\item As a framework to be extended by user-defined objects.
\end{enumerate}
In this section, we show examples of how to solve common problems with pyGSTi in all three modes.  For simplicity's sake, we mostly show 1-qubit examples, and briefly describe the extension to more qubits.  The examples herein were developed using version 0.9.9.1 of pyGSTi.

Some knowledge of Python is assumed.   The examples are valid, but incomplete, so readers should not just paste examples into a Python interpreter as-is.  These snippets present the most salient parts of fully functional examples, leaving out setup steps.  Complete and pedagogical examples are given in the many \href{https://github.com/pyGSTio/pyGSTi/tree/9d516b0b7edf6ff03e75eb148eb58f25c2f34604/jupyter_notebooks/Tutorials}{tutorial notebooks} in the pyGSTi GitHub repository.  A fully functional Python script that implements the examples can be found as an auxiliarly file to this preprint.

%

\subsection{Use Mode 1: pyGSTi as a standalone QCVV tool.}

PyGSTi is often used as a standalone tool for assessing a quantum processor.  In this mode, short snippets of pyGSTi code are invoked from a shell script or Python interpreter.  Data input is via human-readable text files, and results are output either as numbers printed on a terminal, or (more commonly) as a comprehensive HTML or PDF report document.  This ``high-level'' pyGSTi use mode requires very little coding or Python knowledge -- common pyGSTi functions can be invoked as a command-line utility by writing a short shell script.  

\subsubsection{Runing GST on 1-2 qubits and generateing a performance report.\label{sec:um1_gst}}  
A user with access to a QC processor can use pyGSTi to verify that it's working well, predict how accurately it will run specific application circuits, and determine how to tune it up better.  Listing \ref{lst:gst_and_report} below shows how to run GST on a 1-qubit device with idle, $X(\pi/2)$, and $Y(\pi/2)$ operations, and is similar to pyGSTi's \href{https://github.com/pyGSTio/pyGSTi/blob/9d516b0b7edf6ff03e75eb148eb58f25c2f34604/jupyter_notebooks/Tutorials/algorithms/GST-Overview.ipynb}{GST overview tutorial}. 

\begin{python}[caption={Running GST and then generating a report}, label={lst:gst_and_report}, captionpos=b, float=ht]
from pygsti.modelpacks import smq1Q_XYI
exp_design = smq1Q_XYI.get_gst_experiment_design(max_max_length=64)
pygsti.io.write_empty_protocol_data(exp_design, 'ExampleGSTDataDir')

# USER FILLS IN ExampleGSTDataDir/data/dataset.txt and starts new Python session

gst_data = pygsti.io.load_data_from_dir('ExampleGSTDataDir')
gst_protocol = pygsti.protocols.StandardGST('TP,CPTP,Target')
results = gst_protocol.run(gst_data)

report = pygsti.report.construct_standard_report(results, title='GST Example Report')
report.write_html('exampleReportDir')
\end{python}

This example starts by specifying an experiment (collection of circuits) to run on the processor.  PyGSTi can fit data from any set of circuits, in principle, but GST theory stipulates a particular form, and other circuits generally won't provide as much information.  This example leverages a pyGSTi predefined ``model pack'' for this specific gate set, which defines a \texttt{Model} and related lists of \texttt{Circuit}s. Given a maximum circuit depth (\texttt{max\_max\_length}), the model pack can describe a complete GST experiment.  If our processor had a different API (i.e., target gate set), then instead of using \texttt{smq1Q\_XYI} we might need to construct a model from scratch, and use pyGSTi to build a GST experiment for it (see the \href{https://github.com/pyGSTio/pyGSTi/blob/9d516b0b7edf6ff03e75eb148eb58f25c2f34604/jupyter_notebooks/Tutorials/algorithms/advanced/GST-FiducialPairReduction.ipynb}{tutorial} on ``fiducial selection'' and ``germ selection''). Once the experiment is defined, the script writes a template data file that the user must fill with data produced by running these circuits on a processor. 

Next, the script reads that data back from the files, into a data structure (\texttt{gst\_data}) that contains a pyGSTi \texttt{DataSet}. Then, we \emph{run} the GST analysis on this data -- i.e., find a gate set that fits the data -- by invoking a \texttt{StandardGST} protocol object.  In the constructor, we specify three different kinds of estimators (TP, CPTP, and Target), and this produces a \texttt{ModelEstimateResults} object containing three estimated gate sets: 
\begin{enumerate}
\item the TP (trace-preserving) gate set that fits the data best,
\item the best-fit CPTP (completely positive and TP) gate set,
\item the ideal target model (just for comparison).
\end{enumerate}
Finally, \texttt{construct\_standard\_report} creates a \texttt{Report} object capable of creating an interactive, HTML-format report that allows the user to see and visualize an array of error metrics and diagnostics.  Error bars could be added to almost all of those metrics by adding a few lines of code (see the tutorial on \href{https://github.com/pyGSTio/pyGSTi/blob/9d516b0b7edf6ff03e75eb148eb58f25c2f34604/jupyter_notebooks/Tutorials/reporting/ReportGeneration.ipynb}{report generation}).

Running GST on a 2-qubit processor (or subset of a processor) is straightforward.  If the 2-qubit gate set matches a predefined model pack -- say, \texttt{smq2Q\_XYICNOT} -- then all we have to do is replace \texttt{smq1Q\_XYI} with \texttt{smq2Q\_XYICNOT} in listing \ref{lst:gst_and_report}.  However, 2-qubit GST is significantly more resource-intensive; \texttt{gst\_protocol.run} might take several hours on a personal computer.  Fortunately, standard GST is over-complete. Some of this data is redundant and can be elided for efficiency, at the cost of some robustness. A process called ``fiducial pair reduction'' (FPR) finds and removes unnecessary circuits (see the \href{https://github.com/pyGSTio/pyGSTi/blob/9d516b0b7edf6ff03e75eb148eb58f25c2f34604/jupyter_notebooks/Tutorials/algorithms/advanced/GST-FiducialPairReduction.ipynb}{tutorial on FPR}).  Most pyGSTi model packs have precomputed FPRed experiments. We use one of them in Listing \ref{lst:gst_and_2QFPR}, which demonstrates how to run streamlined (FPRed) GST on a 2-qubit processor with the \texttt{smq2Q\_XYICNOT} gate set.  For more information see the pyGSTi \href{https://github.com/pyGSTio/pyGSTi/blob/9d516b0b7edf6ff03e75eb148eb58f25c2f34604/jupyter_notebooks/Examples/2QGST-RunningIt.ipynb}{example on running 2-qubit GST}.

\begin{python}[caption={GST for two qubits using fiducial pair reduction}, label={lst:gst_and_2QFPR}, captionpos=b, float=ht]
from pygsti.modelpacks import smq2Q_XYICNOT
exp_design = smq2Q_XYICNOT.get_gst_experiment_design(max_max_length=16, fpr=True)
pygsti.io.write_empty_protocol_data(exp_design, 'FPR2QGSTData')

# USER FILLS IN FPR2QGSTData/data/dataset.txt and starts new Python session

from mpi4py import MPI
comm = MPI.COMM_WORLD
GB = (1024)**3

gst_data = pygsti.io.load_data_from_dir('FPR2QGSTData', comm=comm)
gst_protocol = pygsti.protocols.StandardGST('TP,CPTP,Target')
results = gst_protocol.run(gst_data, memlimit=5*GB, comm=comm)

report = pygsti.report.construct_standard_report(
             results, title='GST 2Q Example Report w/FPR', comm=comm)
report.write_html('exampleReportDirFPR2Q')
\end{python}

\noindent Aside from changing \texttt{smq1Q\_XYI} to \texttt{smq2Q\_XYICNOT}, the only differences between this and the 1-qubit example are the additional \texttt{fpr=True} argument to \texttt{get\_gst\_experiment\_design}, and the additional \texttt{memlimit} and \texttt{comm} arguments to \texttt{run} (which tell pyGSTi to respect a rough memory limit and to use multiple processors to speed up the analysis).

PyGSTi can also do GST on 3 or more qubits, but this is still an active research area.  No built-in model packs exist yet, so the user must build a model and construct GST sequences directly.  The \texttt{StandardGST} protocol doesn't apply, because GST on 3+ qubits isn't ``standard'' yet.  And this use case requires an alternative report type that doesn't rely on getting a dense representation of $n$-qubit maps.  Listing \ref{lst:gst_4Q} demonstrates GST on a 4-qubit processor.  We emphasize that this is computationally intensive, and recommend using a computer with 10s or 100s of parallel cores.

\begin{python}[caption={GST on 4 qubits}, label={lst:gst_4Q}, captionpos=b, float=ht]
from mpi4py import MPI
comm = MPI.COMM_WORLD

model = pygsti.construction.build_localnoise_model(
    nQubits=4, gate_names=['Gxpi2','Gypi2','Gcnot'],
    availability={'Gcnot': [(0,1),(1,2),(2,3)]}, parameterization='H+S')

singleQfiducials = [(), ('Gxpi2',), ('Gypi2',)]
circuit_struct = pygsti.construction.create_standard_localnoise_sequences(
    nQubits=4, maxLengths=[1,2,4], singleQfiducials=singleQfiducials,
    gate_names=['Gxpi2','Gypi2','Gcnot'], availability={'Gcnot':[(0,1),(1,2),(2,3)]}, comm=comm)
exp_design = pygsti.protocols.StructuredGSTDesign(model, circuit_struct)
pygsti.io.write_empty_protocol_data(exp_design, "4Q_GST")

# USER CREATES DATASET FILE 4Q_GST/data/dataset.txt

data = pygsti.io.load_data_from_dir("4Q_GST")
gst_protocol = pygsti.protocols.GST(model, gaugeopt_suite=None, verbosity=4)
results = gst_protocol.run(data, memlimit=5*GB, comm=comm)

report = pygsti.report.construct_nqnoise_report(results, title="GST 4Q Example Report")
report.write_html("example4QReportDir")
\end{python}

\subsubsection{Running RB to estimate average gate error}

A user who wants a quick and simple way to quantify a quantum processor's overall performance can use pyGSTi to do RB \cite{magesan2011scalable,knill2008randomized,proctor2018direct}.  As for GST, this example starts by defining an experiment, pauses while data are generated from the as-built device, and then analyzes the data.  Listing \ref{lst:rb_and_plot} shows how to perform Clifford (standard) RB \cite{magesan2011scalable} on a  1-qubit processor with the sames gates as in the GST example.

\begin{python}[caption={Clifford randomized benchmarking on 1 qubit}, label={lst:rb_and_plot}, captionpos=b, float=ht]
nQubits = 1
gate_names = ['Gxpi2', 'Gxmpi2', 'Gypi2', 'Gympi2']
pspec = pygsti.obj.ProcessorSpec(nQubits, gate_names, qubit_labels=['Q0'])
depths = [0, 2, 4, 8, 16, 32, 64]  # circuit depths (in Clifford operations) minus 2.
k = 40  # number of random circuits at each length

exp_design = pygsti.protocols.CliffordRBDesign(pspec, depths, k)
pygsti.io.write_empty_protocol_data(exp_design, '1QCliffordRB')

# USER FILLS IN 1QCliffordRB/data/dataset.txt

rb_data = pygsti.io.load_data_from_dir('1QCliffordRB')
rb_protocol = pygsti.protocols.RB()
results = rb_protocol.run(rb_data)
ws.RandomizedBenchmarkingPlot(results)
\end{python}

The experiment is constructed by passing a \texttt{ProcessorSpec} object to \texttt{CliffordRBDesign}, which selects the random circuits and also records metadata about how they are sampled (which is critical for subsequent analysis).  The processor spec specifies how many qubits are available, and the available gates (see the \href{https://github.com/pyGSTio/pyGSTi/blob/9d516b0b7edf6ff03e75eb148eb58f25c2f34604/jupyter_notebooks/Tutorials/objects/advanced/ProcessorSpec.ipynb}{processor-spec tutorial}). This approach makes it easy to do RB on any number of qubits. As in the GST example, the experiment circuits are written to an empty data set file that the user needs to fill in with real data.  After that is accomplished, the script reads in the data (and metadata) with \texttt{load\_data\_from\_dir}, and uses a \texttt{RB} protocol object to fit an RB exponential decay curve to it.  Finally, we visually plot the decay curve and the estimated gate error rate.


Clifford RB on a single qubit is a widely used and implemented method, but pyGSTi's RB module provides some features that are not widely available.  First, pyGSTi can seamlessly generate circuits for RB on 2 or more qubits, which can be tricky (the complexity is in sampling uniformly random $n$-qubit Clifford operators \cite{koenig2014efficiently} and then compiling them into circuits containing only the available gates \cite{hostens2005stabilizer}). Second, pyGSTi's RB workflow is set up to generate circuit lists for running multiple RB experiments together, and to automate analysis of such data.  For example, pyGSTi can easily generate a single circuit list for running 1, 2, 3, 4 and 5-qubit RB, to (e.g.) probe how error rate changes with the number of qubits \cite{proctor2018direct}.  Third, pyGSTi implements two other types of RB -- \emph{direct RB} \cite{proctor2018direct} and \emph{mirror RB} -- that use shallower circuits than Clifford RB and are designed to be feasable on up to around 10 qubits for direct RB and 100 or more for mirror RB.

\begin{python}[caption={Direct randomized benchmarking of each pair of qubits in a 5-qubit ring as well as all 5 qubits.}, label={lst:drb}, captionpos=b, float=ht]
nQubits = 5
qubit_labels = ('Q0', 'Q1', 'Q2', 'Q3', 'Q4')
gate_names = ['Gcnot', 'Gxpi2', 'Gxmpi2', 'Gypi2', 'Gympi2']
connections = [('Q0','Q1'), ('Q1','Q2'), ('Q2','Q3'), ('Q3','Q4'), ('Q4', 'Q0')]
pspec = pygsti.obj.ProcessorSpec(nQubits, gate_names, qubit_labels=qubit_labels,
				availability={'Gcnot':connections})
depths = [0, 2, 4, 8, 16, 32, 64]
k = 40  # number of random circuits at each length

designs = {Qs: pygsti.protocols.DirectRBDesign(pspec, depths, k, qubit_labels=Qs,
           add_default_protocol=True) for Qs in connections + [qubit_labels]}
comb_design = pygsti.protocols.CombinedExperimentDesign(designs)
pygsti.io.write_empty_protocol_data(comb_design, 'AllPairsDirectRB')

# USER FILLS IN AllPairsDirectRB/data/dataset.txt

rb_data = pygsti.io.load_data_from_dir('AllPairsDirectRB')
protocol = pygsti.protocols.DefaultRunner()
results = protocol.run(rb_data)
\end{python}

Listing~\ref{lst:drb} shows how to use direct RB to benchmark a 5-qubit processor with ring connectivity, by benchmarking all 5 qubits together and benchmarking each connected pair of qubits.  As in the previous example, the first step is to create a \texttt{ProcessorSpec} that defines the QC processor and its API.  Then an experiment is created, and a \texttt{CombinedExperimentDesign} writes those circuits to a single template dataset file, using meta-data files to keep track of which circuits correspond to which RB experiment.  Next, the user needs to take data and fill in the (single) template.  Because each experiment design included a default protocol (to run on the data once it was taken), we can run RB on each qubit set by creating and running a single \texttt{DefaultRunner} object on the data. This code can easily be modified to benchmark different qubit subsets, or to use a different distribution of random circuits.

By using \texttt{SimultaneousExperimentDesign} objects, pyGSTi can also perform, with minimal changes to this syntax in Listing~\ref{lst:drb}, a general form of simultaneous RB \cite{gambetta2012characterization,mckay2017three} to probe crosstalk. Further details on running RB with pyGSTi can be found in the \href{https://github.com/pyGSTio/pyGSTi/blob/9d516b0b7edf6ff03e75eb148eb58f25c2f34604/jupyter_notebooks/Tutorials/algorithms/RB-Overview.ipynb}{RB overview tutorial}, which links to in-depth tutorials on the different RB methods available in pyGSTi (some of which are not covered here).

\subsubsection{Testing a model in lieu of running GST}

Sometimes a user already has an error model in mind for their processor, and wants to validate it experimentally.  GST is expensive, and RB doesn't provide enough detail!  This user could use pyGSTi's \emph{model testing} capability to check whether their pre-existing model is consistent with some validation data (e.g., a GST dataset).  Model testing can be especially helpful when the pre-existing model is simple and/or inspired by device physics.  It's also much faster than GST, and can be used on larger processors (up to $\sim 10$ qubits).

Listing \ref{lst:modeltest_and_report} shows how to perform model testing on a single qubit processor with idle, $X(\pi/2)$, and $Y(\pi/2)$ operations, and borrows from the \href{https://github.com/pyGSTio/pyGSTi/blob/9d516b0b7edf6ff03e75eb148eb58f25c2f34604/jupyter_notebooks/Tutorials/algorithms/ModelTesting.ipynb}{pyGSTi tutorial on model testing}.  Because this set of operations coincides with the built-in \texttt{smq1Q\_XYI} model pack, we can follow Listing \ref{lst:gst_and_report} to build a \texttt{gst\_data} object containing a ``target'' \texttt{Model} of the perfect gates and a list of GST circuits.  The example creates and tests a simple depolarizing model (\texttt{model\_to\_test}), but this could be a more complex model based on the device physics (see the tutorials on \href{https://github.com/pyGSTio/pyGSTi/blob/9d516b0b7edf6ff03e75eb148eb58f25c2f34604/jupyter_notebooks/Tutorials/objects/ExplicitModel.ipynb}{explicit} and \href{https://github.com/pyGSTio/pyGSTi/blob/9d516b0b7edf6ff03e75eb148eb58f25c2f34604/jupyter_notebooks/Tutorials/objects/ImplicitModel.ipynb}{implicit} models for details on customizing models in pyGSTi).

\begin{python}[caption={Model testing and report generation}, label={lst:modeltest_and_report}, captionpos=b, float=ht, escapechar=\%]
# define variables from Listing %\ref{lst:gst_and_report}% up until filling in 'data/dataset.txt'

model_to_test = smq1Q_XYI.target_model().depolarize(op_noise=0.07, spam_noise=0.07)
mt_protocol = pygsti.protocols.ModelTest(model_to_test)
results = mt_protocol.run(gst_data)

report = pygsti.report.construct_standard_report(results, title='Model Test Report')
report.write_html('exampleMTReportDir')
\end{python}

\noindent A \texttt{ModelTest} protocol is created, and using the GST circuits and data from Listing \ref{lst:gst_and_report} the \texttt{mt\_protocol.run} function returns a \texttt{ModelEstimateResults} object.  From this we create a HTML report just as in listing \ref{lst:gst_and_report}.  Testing models on 2 or more qubits requires only small changes to the code (analogous to the changes from 1- to 2-qubit GST).

\subsubsection{Testing for drift, and tracking it\label{um1Drift}}
Most QCVV protocols assume \emph{stable} behavior from the QC processor.  But real-world devices often display drift over time, which can corrupt the results of protocols that assume stability \cite{van2013quantum}. Drift can cause visible model violation in GST, and non-exponential decays in RB \cite{fogarty2015nonexponential}.  

PyGSTi implements an end-to-end method for detecting and characterizing drift. Rather than a ``QCVV protocol'' \emph{per se}, this technique is an add-on that can be attached to any other QCVV protocol or quantum circuits experiment to detect and characterize drifting properties by applying statistical tests to \emph{time series} data from quantum circuits \cite{proctor2019detecting}.  It requires each circuit's data to be recorded as a stream of measurement outcomes obtained over time -- e.g., a string of 0s and 1s for a 1-qubit circuit -- rather than a single pooled set of counts.  PyGSTi specifies a simple text format for this data, detailed in the \href{https://github.com/pyGSTio/pyGSTi/blob/9d516b0b7edf6ff03e75eb148eb58f25c2f34604/jupyter_notebooks/Tutorials/objects/advanced/TimestampedDataSets.ipynb}{pyGSTi tutorial on time-stamped data}.  To enable drift detection, the circuits that make up an experiment should be \emph{rastered}, repeating each circuit once and then looping through this sequence repeatedly.  Listing~\ref{lst:drift-report} demonstrates the automated drift analysis.

\begin{python}[caption={Bolting on drift analysis to GST}, label={lst:drift-report}, captionpos=b, float=ht]
from pygsti.modelpacks import smq1Q_XYI
exp_design = smq1Q_XYI.get_gst_experiment_design(max_max_length=16)
pygsti.io.write_empty_protocol_data(exp_design, 'DriftExample')

# USER FILLS IN DriftExample/data/dataset.txt WITH TIME SERIES DATA

gst_data = pygsti.io.load_data_from_dir('DriftExample')
stability_protocol = pygsti.protocols.StabilityAnalysis()
results = stability_protocol.run(gst_data)

report = pygsti.report.construct_drift_report(results, title='GST Drift Report')
report.write_html('DriftReport')
\end{python}

This code snippet generates a drift report, which states whether drift was detected in each circuit.  Where drift was detected, it provides a reconstructed estimate of the circuit probabilities over the duration of the experiment.  Since the experiment in this example was a GST experiment, we could also run the GST analysis on the pooled data.  Then, by cross-referencing the GST report and the drift report, the user could determine how much (if any) of the model violation observed in GST can be explained by drift. Further details can be found in the \href{https://github.com/pyGSTio/pyGSTi/blob/9d516b0b7edf6ff03e75eb148eb58f25c2f34604/jupyter_notebooks/Tutorials/algorithms/DriftCharacterization.ipynb}{drift analysis pyGSTi tutorial}, and Ref.~\cite{proctor2019detecting}.



\subsection{Use mode 2: pyGSTi within a Python analysis toolchain.}

PyGSTi can also be used as a Python package (library), called from a larger analysis procedure written in Python.  This mode is useful for automatically extracting specific information from pyGSTi objects and using it for further analysis.  If Use Mode 1 is ``pyGSTi as an app'', Use Mode 2 is ``pyGSTi as a subroutine''.

\subsubsection{Monitoring a few key gate properties}
A user who wants to regularly monitor a long-running testbed could use pyGSTi's GST capability within a larger analysis script.  They could write a Python script that draws on multiple libraries to:
\begin{enumerate}
\item Trigger at 5pm every Friday evening,
\item Run a predetermined GST experiment on their processor using their own control code,
\item Invoke pyGSTi to process the data and create a \texttt{ModelEstimateResults} object,
\item Extract key error metrics from the \texttt{ModelEstimateResults} object and email them.
\end{enumerate}
In this example, instead of creating a report, we extract values directly from the \texttt{ModelEstimateResults} object.  Listing \ref{lst:using_gst_results} shows how to extract gate matrices from results and compute their infidelity.  It also extracts a quality-of-fit metric using the \texttt{two\_delta\_logl\_nsigma} function (explained below).  This example code could be used for any number of qubits (although it calls \texttt{todense}, which returns a dense process matrix and may cause memory issues if used for too many qubits!).  
\clearpage

\begin{python}[caption={Extracting and computing with gates from GST result}, label={lst:using_gst_results}, captionpos=b, float=ht]
dataset = results.dataset
gst_estimate = results.estimates["CPTP"].models['stdgaugeopt']
ideal_model = results.estimates["CPTP"].models['target']

Gx_matrix = gst_estimate[('Gxpi2',0)].todense()  # numpy array
Gx_idealmatrix = ideal_model[('Gxpi2',0)].todense()  # numpy array

infidelity = pygsti.tools.entanglement_infidelity(Gx_matrix, Gx_idealmatrix)

nSigma = pygsti.tools.two_delta_logl_nsigma(gst_estimate, dataset)
\end{python}


\subsubsection{Interacting with a python-version of the GST report}
The reports that pyGSTi generates in ``standalone'' mode are, by default, in HTML format for interactive viewing through a web browser.  Power users may wish to interact more closely with the results, or extract and manipulate particular bits.

To facilitate this use mode, pyGSTi can also output results in the form of a Jupyter notebook filled with \emph{Python} code that generates all the figures and tables of a standard GST report.  A user can execute parts of this code to generate figures and numbers on demand.  
Listing \ref{lst:notebook_report} shows how to generate a report notebook by replacing the \texttt{write\_html} call seen elsewhere with \texttt{write\_notebook}.
\begin{python}[caption={Generate a report as a Jupyter notebook}, label={lst:notebook_report}, captionpos=b, float=ht]
report = pygsti.report.construct_standard_report(results, title="GST Report Example")
report.write_notebook("myReport.ipynb")
\end{python}

\subsubsection{Exploring benchmarking data}

``Benchmarking'' in pyGSTi refers to protocols that don't fit explicit models to the data.  The canonical example is RB, which just extracts a single error rate.  Power users can easily extract RB error rates from the corresponding results objects and use them in further analysis. For example, error rates from multiple RB experiments can be postprocessed using simple linear algebra to infer (roughly) the error rates of individual gates \cite{proctor2018direct}. 


But ``benchmarking'' in pyGSTi goes beyond RB.  PyGSTi's also contains the \texttt{VolumetricBenchmark} protocol \cite{blume2019volumetric}, which can be run on general benchmarking data.  For example, if RB has been run on $1$\ldots $n$-qubit subsets of a processor, we can extract \emph{volumetric benchmarking} data to explore the processor's capability to run random circuits of various sizes.  For more on volumetric benchmarks within pyGSTi, see the \href{https://github.com/pyGSTio/pyGSTi/blob/9d516b0b7edf6ff03e75eb148eb58f25c2f34604/jupyter_notebooks/Tutorials/algorithms/VolumetricBenchmarks.ipynb}{tutorial}.

\subsubsection{Checking gate calibration with a few simple models}
User scripts can use model testing to rapidly validate models.  For example, suppose that a 2-qubit CNOT gate is known to drift out of calibration periodically.  A user could write a script that periodically runs a small set of user-defined circuits involving CNOT gates, and tests the data against three different models describing (1) a the correct CNOT rotation, (2) an overrotation, and (3) an underrotation.  This custom protocol would efficiently check whether the gate rotation angle had shifted up or down (although a \emph{different} error would void the warranty!).  Listing \ref{lst:modeltest_handpicked} shows such a script, assuming predefined models \texttt{model1}, \texttt{model2}, and \texttt{model3}, user-defined circuits \texttt{circuit\_list}, and externally generated data \texttt{dataset}.

\begin{python}[caption={Choosing the best-fit among several hand-picked models}, label={lst:modeltest_handpicked}, captionpos=b, float=ht]
nSigmas = []
nSigmas.append(pygsti.tools.two_delta_logl_nsigma(
                    model1, dataset, circuit_list))
nSigmas.append(pygsti.tools.two_delta_logl_nsigma(
                    model2, dataset, circuit_list))
nSigmas.append(pygsti.tools.two_delta_logl_nsigma(
                    model3, dataset, circuit_list))
best_model_index = nSigmas.index(min(nSigmas))
\end{python}

This code tests the validity of each model by computing its \emph{likelihood} ($\mathcal{L}$) in light of the available data, and then comparing it to the likelihood of a ``maximal model'' that has enough parameters to fit all the data perfectly. Wilks' theorem\cite{wilks1938large} gives an expected distribution for the difference between the two models' loglikelihoods \emph{if} both are valid.  PyGSTi's \texttt{two\_delta\_logl\_nsigma} function takes a pyGSTi \texttt{Model} and \texttt{DataSet} and returns this number converted into a number of standard deviations.  This measures how confident the user should be that the model is \emph{not} valid -- more standard deviations means higher confidence of detecting a violation.

This code generalizes almost trivially to testing models on multiple qubits.  PyGSTi has common pre-built 1- and 2-qubit models, but operating on 3 or more qubits requires users to build their own models.  Listing \ref{lst:4q_modeltest} shows how to create and test a simple 4-qubit ``local noise'' model where every gate error affects only the gate's target qubit(s).  It is assumed here that \texttt{dataset} contains data for some set of 4-qubit circuits having gates labeled by ``Gxpi2'', ``Gypi2'' and ``Gcnot''.  

\begin{python}[caption={Testing a 4-qubit model}, label={lst:4q_modeltest}, captionpos=b, float=ht]
model_to_test = pygsti.construction.build_localnoise_model(
    nQubits=4, gate_names=['Gxpi2','Gypi2','Gcnot'])

nSigma = pygsti.tools.two_delta_logl_nsigma(model_to_test, dataset)
\end{python}

\subsubsection{Simulating a specific circuit of interest.\label{sec:um2_cicuitsim}}
A well-informed user who already has a model for their noisy processor (obtained from theory, modeling, or GST) may wish to assess the consequences of that noise model for the circuits they want to run.  They can use pyGSTi's circuit simulation subroutines to simulate one or more circuits using that model. Listing \ref{lst:circuitsim} shows how to compute the outcome probabilities of a circuit using the \texttt{Model.probs} method.  The model has four qubits, linear connectivity, CNOT gates between each pair of nearest neighbors, $X(\pi/2)$ and $Y(\pi/2)$ gates on each qubit, and no noise.  A \texttt{Circuit} object is created that specifies the circuit we want to simulate (note the equivalent ways of creating a circuit), and the model's \texttt{probs} function performs the simulation.  The returned \texttt{outcome\_probs} dictionary has keys of the form \texttt{'0001'}, corresponding to the probabilities.  See the \href{https://github.com/pyGSTio/pyGSTi/blob/9d516b0b7edf6ff03e75eb148eb58f25c2f34604/jupyter_notebooks/Tutorials/02-Using-Essential-Objects.ipynb}{``using essential objects'' tutorial} for more information.

\begin{python}[caption={Circuit simulation on a 4-qubit model}, label={lst:circuitsim}, captionpos=b, float=ht]
model = pygsti.construction.build_localnoise_model(
            nQubits=4, gate_names=['Gxpi2','Gypi2','Gcnot'], geometry='line')

c1 = pygsti.obj.Circuit('[Gxpi2:0Gypi2:1][Gcnot:0:1][Gcnot:1:2][Gxpi2:0Gcnot:2:3]@(0,1,2,3)')
c2 = pygsti.obj.Circuit([ [('Gxpi2',0),('Gypi2',1)], ('Gcnot',0,1),
                           ('Gcnot',1,2), [('Gxpi2',0),('Gcnot',2,3)] ], line_labels=(0,1,2,3))
outcome_probs = model.probs(c1)
outcome_probs2 = model.probs(c2)

print(outcome_probs.get('0100',0.0))
\end{python}

\subsubsection{Recalibrating a gate efficiently with RPE}

A user who needs to recalibrate an existing qubit that may have drifted out of spec probably does not want to run full GST -- it's resource-intensive, and measures a lot of quantities that are already known.  To characterize just a single gate rotation angle or \emph{phase}, the user can use the robust phase estimation (RPE) \cite{kimmel2015robust} protocol as implemented in pyGSTi \cite{rudinger2017experimental, meier2019testing}. RPE measures the the phase of a single-qubit gate efficiently and accurately, and does so while remaining robust against a wide variety of SPAM and decoherence errors.



Listing~\ref{lst:rpe} shows how to measure the rotation angle on an approximate $X(\pi/2)$ gate using a noisy $\{X(\pi/2),Y(\pi/2)\}$ gate set.  This code uses one of pyGSTi's built-in ``RPE model packs''.  
The \texttt{RPEModelPack} object's \texttt{get\_rpe\_experiment\_design} takes a maximum length parameter and generates a full RPE experiment with circuits up to that length.  The resulting data are 
analyzed by instantiating an \texttt{RPE} protocol object and running it on a data object (\texttt{rpe\_data}) to produce a \texttt{RPEResults} object containing the estimated rotation angle and ancillary 
debugging information (see the \href{https://github.com/pyGSTio/pyGSTi/blob/9d516b0b7edf6ff03e75eb148eb58f25c2f34604/jupyter_notebooks/Tutorials/algorithms/RobustPhaseEstimation.ipynb}{tutorial on RPE results}).

\begin{python}[caption={Robust phase estimation on 1 qubit}, label={lst:rpe}, captionpos=b, float=h]
from pygsti.modelpacks import smq1Q_Xpi2_rpe
exp_design = smq1Q_Xpi2_rpe.get_rpe_experiment_design(max_max_length=2**6)

pygsti.io.write_empty_protocol_data(exp_design, 'RPEData')

# USER FILLS IN RPEData/data/dataset.txt

rpe_data = pygsti.io.load_data_from_dir('RPEData')
rpe_protocol = pygsti.protocols.RPE()
results = rpe_protocol.run(rpe_data)

print(results)
\end{python}

\subsubsection{Comparing two or more data sets}

The behavior of qubits and QC processors can be altered by a wide variety of external \emph{contexts}, including time of day, temperature, and whether nearby qubits are being driven at the same time.  PyGSTi implements a simple general algorithm to detect such variations (which complements the specific focused algorithms for drift detection that we discussed above).  It compares two nominally identical datasets, uses statistical hypothesis tests to ignore ordinary finite-sample fluctuations between the two, and identifies circuits that have clearly changed between the two runs.  Listing~\ref{lst:dcomp} shows how to use the \texttt{DataComparator} object.

\begin{python}[caption={Comparing datasets that should be the same.}, label={lst:dcomp}, captionpos=b, float=h]
ds1 = pygsti.io.load_dataset("Dataset1.txt")
ds2 = pygsti.io.load_dataset("Dataset2.txt")

comparator = pygsti.objects.DataComparator([ds1, ds2])
comparator.implement()

comparator.get_worst_circuits(10)
\end{python}

This analysis tool is fast, simple, and can be used to investigate many behaviors that constitute \emph{context dependence}. Examples include:
\begin{itemize}
\item Comparing results of GST circuits run while idling all other qubits, to the same circuits run while driving gates on the other qubits.  This detects \emph{crosstalk}. 
\item Comparing circuits run at time $t_1$ to the same circuits run at time $t_2$.  This detects \emph{drift}.  More than one time can also be compared simultaneously. This technique is less powerful than the method of Section \ref{um1Drift}, but easier to implement.
\end{itemize}
For more details see the \href{https://github.com/pyGSTio/pyGSTi/blob/9d516b0b7edf6ff03e75eb148eb58f25c2f34604/jupyter_notebooks/Tutorials/algorithms/DatasetComparison.ipynb}{tutorial on dataset comparison} and Ref.~\cite{rudinger2018probing}.


\subsection{Use mode 3: Customizing pyGSTi for user-specific models and physics.\label{secUseCase3}}
The most sophisticated way that users can interact with pyGSTi is by customizing its models and algorithms to suit the user's specific needs.  In this section, we show how a power user can define custom gate operations or custom models, and then test or fit them against data.  This use mode is supported and encouraged by pyGSTi; it just requires defining a new model and constructing a suitable experiment (set of circuits).  The examples in this section are a bit longer and more technical, because of the high degree of interface between user code and pyGSTi.

\subsubsection{Adding custom constraints to GST\label{sec:um3GST}}

Many device-specific models can be obtained by adding \emph{constraints} to a familiar model.  For instance, if a processor's Z-rotation gate is implemented entirely in software (a phase reference update), a user may want to model this gate as \emph{always} being error-free.  Such a gate has no adjustable parameters.  Listing \ref{lst:gst_custom_param} illustrates how to do GST with such a model for 1 qubit.

There are two basic way to add custom constraints to a model.  The \texttt{set\_all\_parameterizations} method changes the parameterization of (i.e.~the constraints on) \emph{all} the operations within the model.   Alternatively, by assigning an \texttt{Operator} object to a single element of a \texttt{Model}, only the parameterization of that element is changed.  The example code first changes the parameterization of all the operations to being CPTP-constrained, then changes the parameterization of the Z-rotation gate (\texttt{"Gzpi2"} on the one and only qubit 0) to having no parameters (a \emph{static} gate in pyGSTi's language) by replacing it with a \texttt{StaticDenseOp} having the same process matrix.  Finally, GST is performed on the resulting model.  Because the static gate cannot be gauge-optimized, we specify \texttt{gaugeopt\_suite=None}.

\begin{python}[caption={GST using a model with customized a parameterization}, label={lst:gst_custom_param}, captionpos=b, float=ht]
from pygsti.modelpacks import smq1Q_XYZI
initial_model = smq1Q_XYZI.target_model()

# Change all of initial_model's operations to having a CPTP-constrained parameterization.
initial_model.set_all_parameterizations("CPTP")

# Replace the CPTP-constrained Z-gate with an always-perfect (0 parameter, "static") Z-gate.
gate_matrix = initial_model[('Gzpi2',0)].todense()
initial_model[('Gzpi2',0)] = pygsti.objects.StaticDenseOp(gate_matrix)

# Run GST (on existing gst_data)
results = pygsti.protocols.GST(initial_model, gaugeopt_suite=None).run(gst_data)
\end{python}


\subsubsection{Running GST using a custom gate parameterization\label{sec:um3CustomGate}}
Hardware designers often have a specific physics-inspired model for how a gate might vary, which has only a few adjustable parameters.  Physics-based models constrain the types of errors that need to be modeled and measured.  Listing \ref{lst:custom_operator} demonstrates how to create a user-defined \texttt{Operator} object, taken from pyGSTi's \href{https://github.com/pyGSTio/pyGSTi/blob/9d516b0b7edf6ff03e75eb148eb58f25c2f34604/jupyter_notebooks/Tutorials/objects/advanced/CustomOperator.ipynb}{custom operator tutorial}.  A new class is defined that inherits from \texttt{DenseOperator}, which represents an operator that internally stores a dense process matrix representation of itself.  The workhorse function is \texttt{from\_vector}, which takes in a vector of two (since the operator declares that it has 2 parameters in \texttt{num\_params}) real values and produces a process matrix.  This example creates a single-qubit $X(\pi/2)$ gate parameterized by a depolarization rate and an over-rotation.  (If both parameters are zero, then the gate is a perfect $X(\pi/2)$ rotation.)

\begin{python}[caption={Defining a custom operator object}, label={lst:custom_operator}, captionpos=b, float=ht]
class MyXPi2Operator(pygsti.obj.DenseOperator):
    def __init__(self):
        # initialize with no noise
        super().__init__(np.identity(4,'d'), "densitymx")
        self.from_vector([0, 0])

    def num_params(self):
        return 2  # we have two parameters

    def to_vector(self):
        return np.array([self.depol_amt, self.over_rotation], 'd')

    def from_vector(self, v, close=False, nodirty=False):
        # initialize from parameter vector v
        self.depol_amt = v[0]
        self.over_rotation = v[1]

        theta = (np.pi/2 + self.over_rotation)/2
        a = 1.0-self.depol_amt
        b = a*2*np.cos(theta)*np.sin(theta)
        c = a*(np.sin(theta)**2 - np.cos(theta)**2)

        # .base is a member of DenseOperator and is a numpy array that is
        # the dense Pauli transfer matrix of this operator
        self.base[:,:] = np.array([[1,   0,   0,   0],
                                   [0,   a,   0,   0],
                                   [0,   0,   c,  -b],
                                   [0,   0,   b,   c]], 'd')
        if not nodirty: self.dirty = True

    def transform(self, S):
        # Update self with inverse(S) * self * S (used in gauge optimization)
        raise NotImplementedError("MyXPi2Operator cannot be transformed!")
\end{python}

Objects of type \texttt{MyXPi2Operator} can then be instantiated and added to a model, and running GST proceeds in the usual way, shown in listing \ref{lst:gst_custom_operator}.  A technical difference is that, because the custom gate doesn't possess a clearly defined gauge group, models containing it cannot be ``gauge-optimized'' in the way standard GST models are.  This is again specified by setting \texttt{gaugeopt\_suite=None} in the code.

\begin{python}[caption={Performing GST with a model containing a custom operator}, label={lst:gst_custom_operator}, captionpos=b, float=ht]
from pygsti.modelpacks import smq1Q_XYI
model = smq1Q_XYI.target_model()
model.operations[('Gxpi2',0)] = MyXPi2Operator()

# Run GST *without* gauge optimization
gst_protocol = pygsti.protocols.GST(model, gaugeopt_suite=None)
results = gst_protocol.run(gst_data)
\end{python}

\end{widetext}

\section{pyGSTi as a framework\label{secFramework}}
As shown above, pyGSTi can perform a lot of QCVV tasks out of the box.  But Use Case 3 hints at another perspective:  pyGSTi is really a flexible framework for doing certain things, which makes it easy to build new tools on the fly.  In this section we address the question ``What does pyGSTi do?'' from a slightly new angle, presenting pyGSTi as providing generic functionality useful for building other software.

At heart, pyGSTi is a general framework for constructing, testing, optimizing, and visualizing models given data.  Its capability to construct almost \emph{any} QCVV model and compare it with almost \emph{any} data is a powerful capability in its own right!  The GST and model-testing examples listed above are just special cases of this general capability.  Here are some important broad capabilities that pyGSTi presents.

\begin{enumerate}
\item \textbf{Defining a noisy-processor model for a device.}  Within pyGSTi, it's possible to describe the behavior of a (potentially) noisy quantum processor in a variety of ways.  Users can provide the types and rates of certain errors, or construct raw process matrices that describe the action of noise.  They can define their own local-gate classes (see section \ref{secUseCase3}), to easily incorporate customized gate parameterizations into a model.  In multi-qubit contexts, noise parameters can be modeled as affecting one, a few, or all of the available qubits.  PyGSTi's models allow great freedom in describing how ``layers'' -- $n$-qubit CPTP maps representing all the gates in a single clock cycle -- are constructed from lower-qubit operations, and these construction rules are designed to be customized.  In this way pyGSTi can easily be adapted to use physical/microscopic device models, and is not limited to just few-qubit models.  Gates and models can depend explicitly on time.  The pyGSTi API has been designed from the ground up as a framework in which almost arbitrary \emph{parameterized} models can be created and manipulated.  
\item \textbf{Finding experiments (circuits) that probe a model.}  After a model is created, pyGSTi contains functionality for finding a list of circuits that probe all of the parameters of that model.  This is a necessary pre-processing step that automates the process of deciding \emph{how} to accurately estimate the adjustable parameters within a model (i.e., what circuits should be run).  The algorithms that perform this experiment design (or ``circuit selection'') are specific to GST-like characterization and are not implemented by many, if any, other software packages.
\item \textbf{Fitting or testing a model to experimental data.}  PyGSTi has routines for optimizing the likelihood between a model and circuit-outcome data (possibly time-resolved data).  While at its core this is just a run-of-the-mill optimization routine adapted to working with pyGSTi's model and data set objects, nontrivial modifications to the algorithm have been made to significantly increase its performance and allow it to be used on multiple processors.
\item \textbf{Visualizing and interpretating models.}  Noisy-processor models can have just a few adjustable parameters, or thousands of them.  Models with few parameters are usually easy to interpret, because they were constructed with a particular physical interpretation in mind.  Ones with many adjustable parameters are usually much harder to interpret, because they can describe such a wide range of underlying physical phenomena.  GST normally uses ``big'' models, and pyGSTi contains functionality for interpreting models that represent each gate as a mostly-arbitrary process matrix.  In particular, pyGSTi has extensive routines for dealing with non-physical ``gauge'' degrees of freedom -- e.g., ``gauge-optimization'' routines that find sensible representations (from among a large class of physically equivalent ones) of noisy gates.  PyGSTi contains general functionality for identifying and defining gauge degrees of freedom. 
\item \textbf{Displaying lots of data interactively.}  PyGSTi contains a framework for creating plots and tables using Javascript and HTML.  This user-interface framework is used to place the analysis information coming out of pyGSTi within web pages and Jupyter notebooks, which we have found to be convenient platforms for many users.  Importantly, the HTML/Javascript format allows \emph{interactivity}, so plots and tables can hold more information than they display at any one time.  PyGSTi contains a common system for generating HTML reports \emph{and} interactively displaying plots and tables within a user's Jupyter notebook.  Thus, the process of creating customized reports requires little more than taking the commands from a user's Jupyter notebook and adding a HTML template.  While the design of this interactive-output capability was again motivated by the large models and correspondingly rich output of GST, these capabilities can prove useful in many other contexts as well.
\end{enumerate}

\section{The big picture:  lessons learned and design insights}

In most of this paper, we've addressed ``What does/can pyGSTi do?'',  focusing on concrete tasks and coding frameworks.  But pyGSTi has been continuously under development since 2013, and in building it to where it is now, we learned a few lessons that may be useful to other researchers.

One is simply that \emph{fitting arbitrary models to data requires tricks.}  From our prior experience with maximum likelihood state and process tomography, we naively thought that the optimization routines from \texttt{Scipy.optimize} would meet our needs.  They didn't!  Even relatively canonical optimizations encountered in QCVV -- like $L_1$-regularized tomography -- cause problems for canned code.  The more general cases that pyGSTi deals with involve highly nonconvex loglikelihood functions, gauge and nuisance parameters, and nonholonomic constraints.  The custom optimization routines now in pyGSTi outperform any others we've encountered \emph{for these specific problems}.  Sometimes, even the best generic optimizer isn't good enough.

We also didn't expect forward simulation of circuits to be nontrivial.  You just multiply process matrices, right?  But we learned (and it sounds obvious in retrospect) that during the course of model optimization the \emph{same} circuits are simulated \emph{many} times.  This changes the computational profile in ways that we didn't initially appreciate.  Today, pyGSTi incorporates elaborate caching mechanisms within its circuit simulation routines that optimize them for the problem of simulating the same (large) set of circuits many times.  Most circuit simulators are not optimized for this mode, and our optimizations produced speedups of roughly 10-100x (this figure depends on many things).  This is a big deal -- it's the difference between a 15-minute analysis and a 24 \emph{hour} one.

Finally, we have learned to fear, loathe, and respect gauge degrees of freedom.  The presence of nonphysical degrees of freedom in the most common and most computationally convenient ways of representing a model within computer memory has been a recurring complication, both for model optimization, and for presenting and interpreting results.  Gauge degrees of freedom make Jacobians and Hessians rank deficient, which confuses many optimizers.  When presenting results, we find that common gauge-\emph{variant} metrics (e.g. fidelity and diamond norm) are very sensitive to the gauge freedoms.  Even a slight change in gauge (which has \emph{no} effect on any physical observable!) can change fidelities by 10x.  We are now actively researching metrics that don't depend on gauge. But for the mean time, computing reasonable fidelities and diamond norms remains essential, and fine-grained QCVV protocols need to deal with gauge freedom.

The authors plan to continue to improve and augment pyGSTi -- both to support our own research, and to serve the wider QC community.  We hope others will continue to find it useful, and to solve problems with it.

Sandia National Laboratories is a multimission laboratory managed and operated by National Technology \& Engineering Solutions of Sandia, LLC, a wholly owned subsidiary of Honeywell International Inc., for the U.S. Department of Energy's National Nuclear Security Administration under contract DE-NA0003525.  This paper describes objective technical results and analysis. Any subjective views or opinions that might be expressed in the paper do not necessarily represent the views of the U.S. Department of Energy or the United States Government.

\bibliography{Bibliography}

\appendix
\begin{widetext}

\section{Screenshots from pyGSTi reports}
Below are several screenshots from an example GST report.  These HTML reports are viewed within a web browser, and display interactive content (e.g. data is displayed as mouse is hovered over different areas of a plot).
\begin{figure}
  \begin{center}
    \includegraphics[width=5.8in]{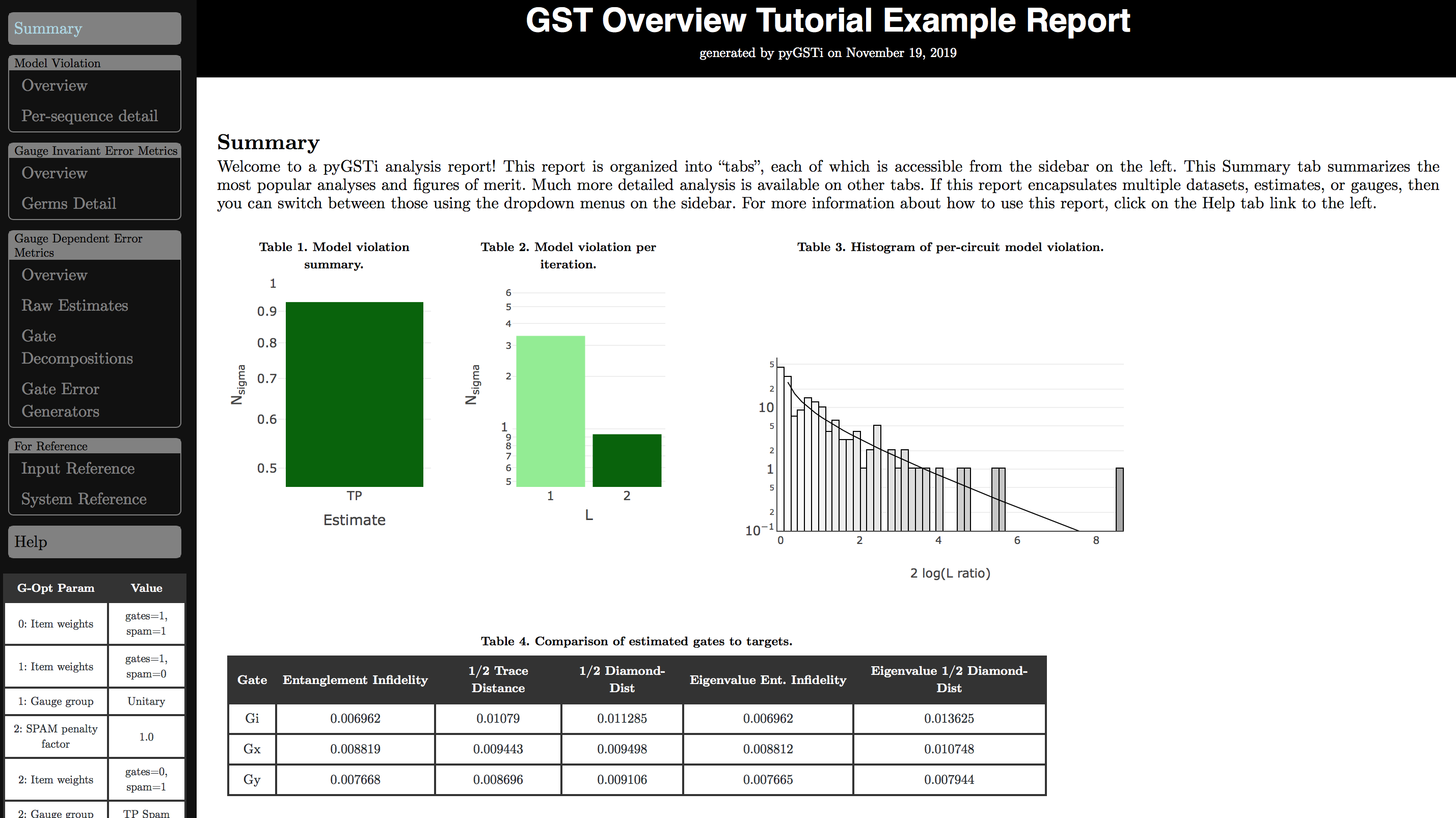}
    \caption{The summary page of a pyGSTi GST report.  Navigation links on the left sidebar move the viewer between different ``tabs'' of the report, each giving different information.  Green bars indicate that the best-fit model describes the data well, and a table of the most common per-gate metrics is given.}
  \end{center}
\end{figure}

\begin{figure}
  \begin{center}
    \includegraphics[width=5.8in]{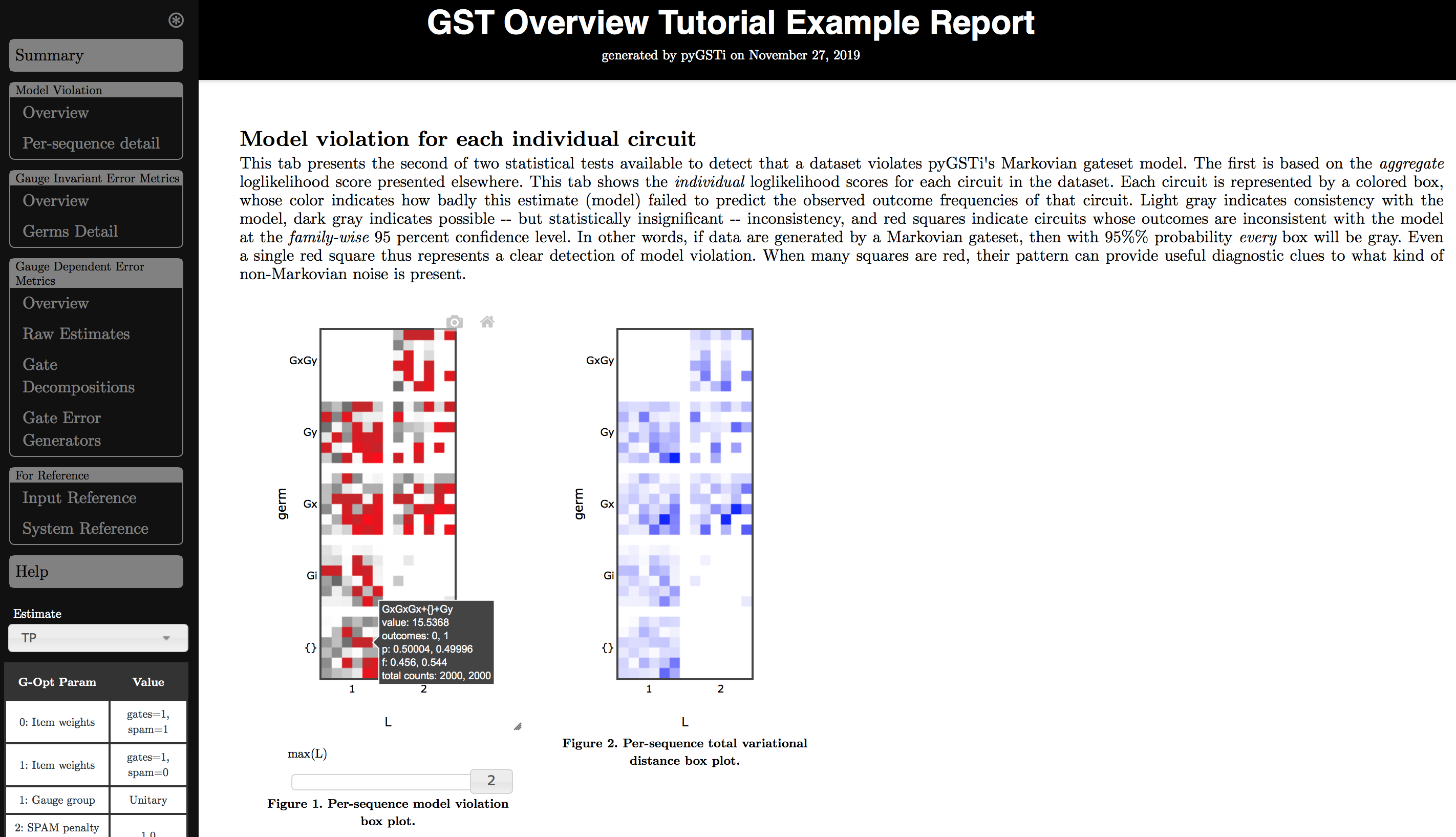}
    \caption{The ``per-sequence detail'' tab of another (different) example GST report where the best-fit model could \emph{not} predict the data so well.  Squares of colored boxes show how well GST's best-fit model predicted the outcomes of each circuit.  The grayscale transitions to a redscale at the point where deviations become statistically significant.  Hovering over a box shows the predicted probability and observed frequency for each outcome of that circuit (the image shows this for the \texttt{"GxGxGxGy"} circuit).}
  \end{center}
\end{figure}
\end{widetext}
\end{document}